# Integrated photonic building blocks for next-generation astronomical instrumentation II: the multimode to single mode transition

**Izabela Spaleniak,**[1,2*] **Nemanja Jovanovic,**[3] **Simon Gross,**[1,5] **Michael J. Ireland,**[1,2,4] **Jon S. Lawrence,**[1,2,4] **and Michael J. Withford**[1,2,5]

[1]*MQ Photonics research centre, Dept. of Physics and Astronomy, Macquarie University, NSW 2109, Australia*
[2]*Macquarie University Research Centre in Astronomy, Astrophysics & Astrophotonics, Dept. of Physics and Astronomy, Macquarie University, NSW 2109, Australia*
[3]*National Astronomical Observatory of Japan, Subaru Telescope, 650 N. A'Ohoku Place Hilo, Hawaii 96720, U.S.A*
[4]*Australian Astronomical Observatory (AAO), 105 Delhi Rd, North Ryde NSW 2113, Australia*
[5]*Centre for Ultrahigh Bandwidth Devices for Optical Systems (CUDOS)*
[*]*izabela.spaleniak@mq.edu.au*

**Abstract:** There are numerous advantages to exploiting diffraction-limited instrumentation at astronomical observatories, which include smaller footprints, less mechanical and thermal instabilities and high levels of performance. To realize such instrumentation it is imperative to convert the atmospheric seeing-limited signal that is captured by the telescope into a diffraction-limited signal. This process can be achieved photonically by using a mode reformatting device known as a photonic lantern that performs a multimode to single-mode transition. With the aim of developing an optimized integrated photonic lantern, we undertook a systematic parameter scan of devices fabricated by the femtosecond laser direct-write technique. The devices were designed for operation around 1.55 μm. The devices showed (coupling and transition) losses of less than 5% for $F/\# \geq 12$ injection and the total device throughput (including substrate absorption) as high as 75-80%. Such devices show great promise for future use in astronomy.

## 1. Introduction

Over the past three decades there has been an increasing trend of exploiting photonic technologies for astronomical instrumentation [1]. This began with multimode optical fibers used for transporting light from the telescope to remote instrumentation, through laser frequency combs for precise wavelength calibration [2] up to the recent application of photonic chips for multi-aperture interferometry [3], pupil remapping interferometry [4], and spectroscopy [5,6].

Recent work has focused on the development of a single mode spectrograph [7-10]. Such an instrument has several key advantages over traditional (multi-moded) spectrographs: precision, small footprint, and a relatively low cost. One of the key features of single-mode fibers is the stable mode profile, meaning that the field distribution at the output of a single mode fiber is independent of changes to the input wavefront. The high stability of the point spread function (PSF) profile is a crucial factor for performing high accuracy spectroscopic measurements, e.g. radial velocity measurements, which are currently limited by modal-noise from multimode fibers. The size of the optical components in the spectrograph near Littrow configuration ($D_{pupil}$ – pupil diameter) is given by the formula:

$$D_{pupil} = \frac{R \cdot \Delta\theta_{slit} \cdot D_{telescope}}{2 \cdot tan\theta_B \cdot \lambda} ,  \quad (1)$$

where $R$ is the spectral resolving power ($\lambda/\Delta\lambda$), $\theta_{slit}$ is the angular slit width, $D_{telescope}$ is the telescope diameter, $\theta_B$ is the blaze angle of the grating, and $\lambda$ is the central wavelength. For a given telescope ($D_{telescope}$) and parameters of the spectrograph ($\lambda$, $R$) the parameters which determine the size of the optics are the slit width ($\Delta\theta_{slit}$) and the blaze angle of the grating ($\theta_B$). There is a limit on $\theta_B$, i.e. in practice it should not be more that ~70°, therefore the main parameter which determines $D_{pupil}$ is $\theta_{Slit}$. Hence the smaller the slit width, the smaller $D_{pupil}$ which then reduces the size of all optical components of the instrument and thereby its footprint. The slit size should be minimized while attempting to capture all of the light at the focus of the telescope. However, when observing from the ground using a moderate sized telescope ($D_{telescope} > 1$ m), where the Earth's atmosphere perturbs the wavefront of the signal, the PSF of the stellar target is highly distorted and surrounded by hundreds of diffraction spots (speckles), each of them formed by coherent wavefront patches. The timescale of this process is very short (< 10 ms), meaning that the image integrated over even as little as 1 s results in a cloud, spatially dispersed pattern that is a superposition of many speckles (known as a seeing pattern) which is well approximated by a Gaussian distribution [11]. In order to utilize all the light in this regime, the spectrograph slit should be of the size of the full width at half maximum (FWHM) of the seeing spot. As the seeing spot is at least one order of magnitude larger than the diffraction limit of the telescope (e.g. 1.2-m UK Schmidt Telescope at the Siding Spring Observatory: $\theta_{diffraction}$=0.12" and $\theta_{seeing}$=1.20" [12]) then the slit size must be increased along with the size of the spectrograph.

In the past ~10 years, the atmospheric effects have been ameliorated using adaptive optics (AO) systems which analyze the incoming corrugated wavefront and correct it using a deformable mirror in real time. If the correction is good enough, the signal can be then focused into a quasi-diffraction-limited spot which enables efficient coupling to single-mode fibers. The major drawbacks of AO systems are: high cost, the need for bright guide stars and poor performance in the visible wavelength range. The poor visible performance results from the shorter coherence times and smaller coherence lengths of the atmosphere in the visible compared to the infrared, and because the atmosphere cannot be well-approximated by a thin phase-screen, requiring multi-conjugate adaptive optics.

An alternative solution to mitigate the disruptive atmospheric effects is to convert a seeing-limited PSF into a diffraction-limited spot or, in other words, to convert a multimode signal into a single-mode signal. The transition of $N$ modes into one mode would violate the second law of thermodynamics, therefore the multimode (MM) signal needs to be reformatted into multiple single-mode (SM) signals. The device which can realize this process is called a *photonic lantern*.

The *photonic lantern* is a mode reformatting device [13]. The concept was first proposed by Leon-Saval *et al.* in 2005 [14]. They showed that a highly efficient transition is possible providing that the number of SM fibers matches the number of spatial modes in the multimode fiber. Simultaneously, the MM end of the photonic lantern needs to support all the modes in the focus of the telescope. The number of modes $N_m$ (for each polarization) in the focus of a telescope is given by [15]:

$$N_m \approx \frac{V^2}{4} = \frac{\pi^2}{16}\left(\frac{\theta_{seeing} \cdot D_{telescope}}{\lambda}\right)^2, \qquad (2)$$

where $V$ is the normalized frequency, $\theta_{seeing}$ is the seeing of the site (in rad), $D_{telescope}$ is the telescope diameter and $\lambda$ is the free space wavelength. For $D_{telescope}= 0.4$ m, at a site with $\theta_{seeing} = 2$" the number of modes at $\lambda =1.55$ μm is $N_m \approx 4$, whereas at $\lambda = 0.45$ μm is $N_m \approx 49$. This means that the number of modes increases dramatically with decreasing wavelength. The first efficient photonic lanterns were demonstrated by Noordegraaf *et al*. [16]. Their device was fabricated using seven single-mode optical fibers fused together and tapered down to create a multimode end. It was designed for efficient operation at 1.55 μm.

The initial motivation for these prototype photonic lanterns was to build a device that would suppress the telluric emission from OH lines in the atmosphere using fiber Bragg gratings [17-18]. Fiber Bragg gratings, like most photonic devices work best with single-mode fibers, therefore photonic lanterns were the ideal link to the multimode collection fiber at focus. The GNOSIS instrument uses seven pairs of photonic lanterns back-to-back (MM-SM-MM), which are connected to complex fiber Bragg gratings in the middle that suppress 103 atmospheric lines between 1.47-1.7 μm [19-20].

Although the fiber based photonic lanterns offer high performance, they are limited in terms of scalability, they are bulky, temperature sensitive, and complex to make and handle. Therefore it was proposed [21] that ultrafast laser inscription could be used to fabricate three-dimensional (3D) photonic lanterns in a monolithic device. In this technique ultrashort laser pulses are focused in a dielectric material like glass. If sufficient power is delivered the chemical structure of the material can be locally modified which results in a permanent refractive index change [22]. By translating the glass sample along three axes, a track of index change can be inscribed which acts as an optical waveguide. The waveguides can in principle be packed arbitrarily close, which allows for the miniaturization of devices compared to fiber-based technology. Thomson *et al*. [23] fabricated the first integrated photonic lanterns that supported 16 modes at 1.55 μm. The MM-SM-MM devices exhibited too high losses to be used in astrophotonics (5.7 dB of insertion loss, throughput of 27%), but they showed that the loss for a MM-SM transition could be ~2.0 dB (63% throughput) and therefore they demonstrated the feasibility of the approach.

In our previous study [24] we demonstrated efficient MM guides for collecting the light from the focal plane of a telescope. In this body of work we build on that and the work by Thomson *et al*. [23] by outlining the procedure for obtaining optimized integrated photonic lanterns. In section 2 we outline the design and simulation of the devices. In section 3 we describe the fabrication and characterization of the integrated lanterns while section 4 highlights their performance. The performance of a slit reformatting device is presented in section 5 along before some concluding remarks.

## 2. Modeling photonic lanterns

### 2.1 Photonic lantern design

The basic components of a photonic lantern include the MM waveguide for collecting the light from the telescope, the MM to SM transition and the isolated SM waveguides as shown in Fig. 2(a) and Fig. 3(a). In our previous study [24], we demonstrated that efficient MM guides could be constructed from a lattice of SM guides in close proximity to each other. The lowest loss devices consisted of 19 SM waveguides arranged in a circular lattice with a 10 μm pitch creating a MM section which was ~50 μm in diameter (Fig. 3(a)). The key feature of these guides is that the SM tracks are slightly smaller than the pitch so that they do not overlap but are sufficiently close to allow for significant cross-coupling. At the MM end, for this design a maximum of 12 non-degenerate modes can be supported (i.e. when fed with a beam focal ratio equivalent to the numerical aperture). In our new photonic lantern device design at the SM end the waveguides were separated by 37.5 µm, so that cross-coupling is eliminated [25].

The main difference between our approach as compared to fiber based lanterns [16], is that we maintain the physical widths of the SM tracks and only bring them closer during the transition while in fiber based devices the cores are tapered in size as well. Although we do not taper the widths it is possible to do this using ultrafast laser inscription by simply varying the pulse energy and/or translation speed as the sample is moved through the focus which adds an extra degree of freedom for designing the lanterns. As the SM tracks do not overlap, there are no regions that will be modified by the laser multiple times and hence this will minimize scattering and radiation losses due non-uniformity in the index profile. Thomson *et al.* [23] used the multiscan technique to fabricate the integrated photonic lanterns. The technique requires multiple scanning of the glass sample back and forth in order to introduce a high enough refractive index modification with small side offsets in order to scan over the volume required. The written structures were of a square shape, which contributes to extra coupling losses from a circular MM input fiber. The MM inputs of the photonic lanterns presented in this work are of a circular shape matching the diameter of a standard 50 µm core diameter MM fiber.

As outlined in the introduction a key requirement for an efficient conversion from a MM input to a SM format is that the number of SM guides need to be equal to or greater than the number of non-degenerate modes in the MM section. For devices consisting of MM-SM-MM transitions, the number of modes supported by the second MM section also has to be equal to or greater than the number of SM guides. Therefore ideally the number of non-degenerate modes in each of the MM sections and the number of SM guides are matched. In general, the maximum number of modes supported by the MM section depends on the size of the waveguides: it increases quadratically with the effective waveguide diameter [15], and the device NA ($N_m \approx \pi^2 d^2 NA^2 / (4\lambda^2)$), where $d$ is the waveguide diameter). RSoft FemSIM [26] was used to calculate the exact number of modes supported by the circular lattice geometry of 19 straight waveguides as a function of pitch. The simulation results are presented in Fig. 1. Up to a pitch of 12 µm, the guides support ≤ 13 non-degenerate modes, at a pitch of 13 µm they support 17 modes and then 19 modes for a pitch of ≥ 14 µm. There are two consequences arising from this result. Firstly, the first MM-SM transition should be very efficient as the number of non-degenerate modes in the MM section is always less than or equal to the number of SM tracks (19). Secondly, the back SM-MM transition is lossy when the number of SM tracks is greater than the number of non-degenerate modes in the second MM section, which is the case for a MM section with a pitch of ≤ 13 µm. Nevertheless, we chose the pitch of 10 µm, as it proved to be the most efficient in the previous study [24].

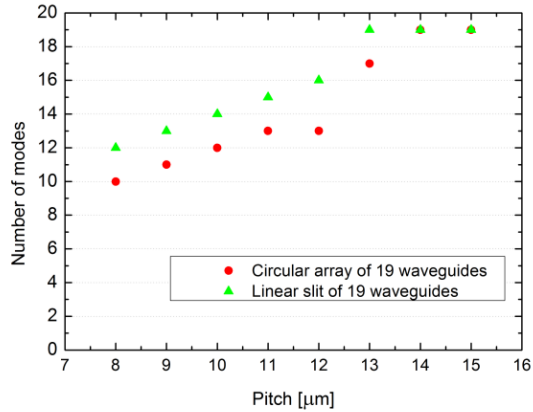

Fig. 1. Number of modes supported by the circular lattice structure of 19 waveguides and a linear slit as a function of the pitch.

*2.2 Beam propagation modeling of transitions*

A commercial implementation of the beam propagation algorithm (BeamPROP package of RSoft [26]) was used to simulate the photonic lanterns, which are conceptually presented in Figure 2(a). A refractive index profile of a typical laser-inscribed SM track was taken with a refractive index profilometer (Rinck Elektronik), and the data inserted into the program for each SM track. The index profile was rescaled to compensate for material dispersion between 0.633 μm (where the index profile was measured) and 1.55 μm (where the guides operate) by using the Sellmeier equation for Corning Eagle2000 (the glass of choice for experiments). A MM launch field was used for the simulations which had a step index profile. We have simulated the behavior for three types of transition: linear, cosine and raised sine ($\mathrm{rsin}(z)=z-\sin(2\pi z)/2\pi$) for transition lengths varying between 1 and 24 mm.

Figure 2(b) presents the simulation results for a wavelength of 1.55 μm. The figure shows the normalized throughput as a function of transition length for all three transition types. For taper lengths between 1-5 mm, the losses strongly depend on the taper length. For transition lengths > 5 mm the transmission reaches ~0.97 and flattens off. This high throughput is due to the fact that the MM section supports 12 modes whereas there are 19 SMs, as mentioned in section 2.1 The cosine type of transition offered the best performance for short transition lengths, therefore it was chosen for all fabricated devices presented in this work.

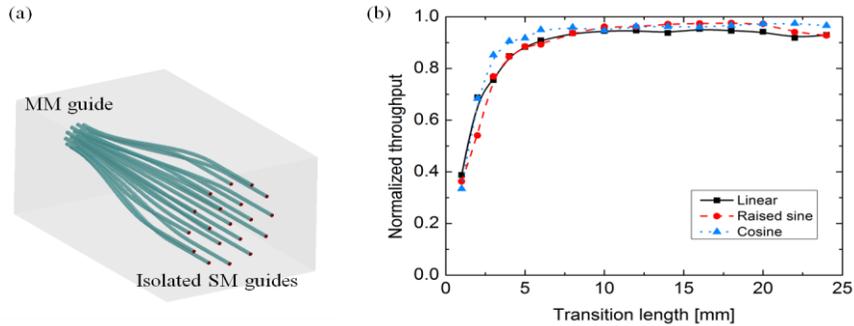

Fig. 2. a) CAD model of a photonic lantern (MM-SM) used in modeling; b) Simulated normalized throughput as a function of transition length between MM and SM sections for three different transition types: linear, raised sine and cosine.

## 3. Waveguide fabrication and characterization

The photonic lanterns were inscribed using an ultrafast Ti:sapphire oscillator (FEMTOSOURCE XL 500, Femtolasers GmbH), which emits pulses at 800 nm center wavelength with 5.1 MHz repetition rate and a pulse duration of < 50 fs. The laser was focused into a 30 mm long alkine earth boro-aluminosilicate (Corning Eagle2000) glass sample using a 100× oil immersion objective lens (Zeiss N-Achroplan, numerical aperture (*NA*) = 1.25, working distance = 450 μm). The sample was placed on a set of Aerotech, 3-axis air-bearing translation stages to ensure smooth translation during writing.

The waveguides were written with a pulse energy of 35 nJ at a translation velocity of 750 mm/minute in order to create waveguides which were single-mode at 1.55 μm. The translation velocity was selected based on our previous study [24], which showed 750 mm/minute to be optimum for MM structures. The chip was ground and polished to reveal the waveguide ends, which resulted in a final sample length of 28.3 mm. As a result of the high repetition rates used all waveguides presented herein were written in the cumulative heating regime [27].

In order to find the optimum design for the photonic lantern the following sets of photonic lanterns were written: A) four sets of back-to-back photonic lanterns (MM-SM-MM) with different transition lengths (1 - 10 mm) as shown in Fig. 3(b), B) nine sets of MM-SM-

MM photonic lanterns with different MM pitches (8 - 15 μm) seen in Fig. 3(b) and C) five sets of MM-SM slit reformatting devices with different slit pitches (7-15 μm) seen in Fig. 3(c).

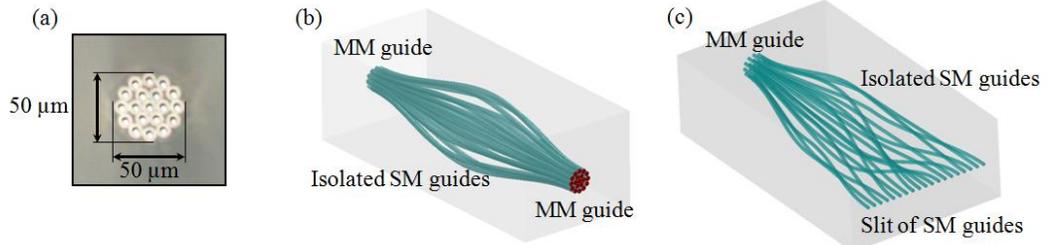

Fig. 3. (a) Microscopic image of the MM input of the photonic lantern and CAD models of (b) the back-to-back photonic lantern (MM-SM-MM) and (c) slit reformatting device (MM-SM).

The photonic lanterns were characterized using the setup depicted in Fig. 4. An IR light emitting diode (LED) with a center wavelength of 1.55 μm and a FWHM bandwidth of 0.115 μm was used as the probe source (Thorlabs-LED1550E). It was driven by a 60 mA current source. The light was directly coupled into a 400 μm core diameter MM fiber and the output of the fiber was collimated with a 50 mm focal length achromat (Thorlabs-AC127-050-C). In order to control the focal ratio $F/\# $ ($\approx 1/(2 \times NA)$) of the injected beam a calibrated iris (Thorlabs-SM1D12C) was placed in the collimated beam. A 20 mm focal length lens was used to focus the light into the sample. The sample was placed on a XYZ precision translation stage (Newport, M-462-XYZ-M). In order to align the sample with the injecting beam a cube beamsplitter (CM1-BS015) was placed in the collimated beam, so that a camera imaged the input facet of the glass sample. The IR LED was temporarily replaced with a green LED and the sample aligned until the image of the input beam (green spot) overlapped with the photonic lantern input. In order to perform throughput measurements the near-field output of the photonic lanterns was reimaged onto an IR camera (FLIR SC7000). The normalization was done by comparing near-field images of the output of the photonic lantern (MM output) and the input beam itself. For this step the photonic chip was removed and the reimaging lens was moved toward the injection lens by the length of the chip. The size of the input beam (Fig. 5) was larger than the size of the MM input, therefore in order to calculate the correct input signal, a virtual aperture matching the input of the MM waveguide (~54 μm) was applied to the input beam in post-processing, which accounted for coupling due to the spot size mismatch.

As a final step, the normalized throughput was rescaled to remove the effect of absorption by the Eagle2000 substrate, so that the true performance of the photonic lanterns could be assessed. Based on the 0.0065 mm$^{-1}$ absorption coefficient in Eagle2000 at 1.55 μm [28], the maximum possible throughput for the 28.3 mm long sample was limited to 83.0%. All throughputs reported hereafter have been rescaled to compensate for this.

Measurements of the $F/\#$ of the launch was carried out by using the assembly depicted in the top right of Fig. 4. The IR camera imaging lens was focused onto a plane 1 focal length behind the reimaging lens (its pupil plane), creating an image on the detector which represents the far-field of the probed beam. Dark frames were subtracted from the data initially and then the 90% encircled energy spot size was calculated for each image. The image scale was calibrated by means of a ruler placed in the same plane as the reimaging lens. The *NA* was calculated from the following equation; *NA = spot size on the detector/(2×focal length of the reimaging lens)*. The *F/#* was determined by then using *F/# ≈1/(2×NA)*.

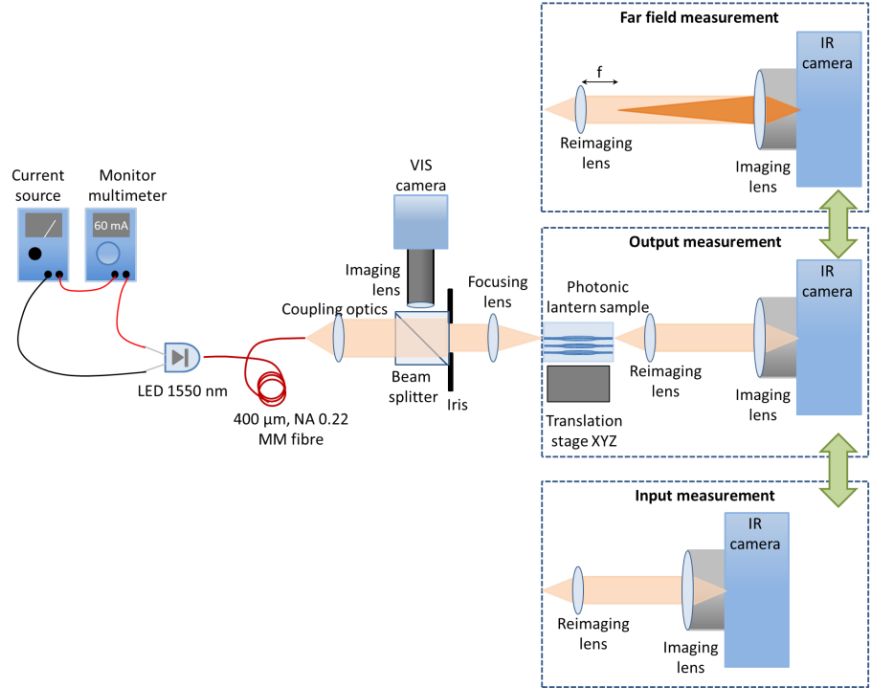

Fig. 4. Schematic diagram of setup for characterizing photonic lanterns.

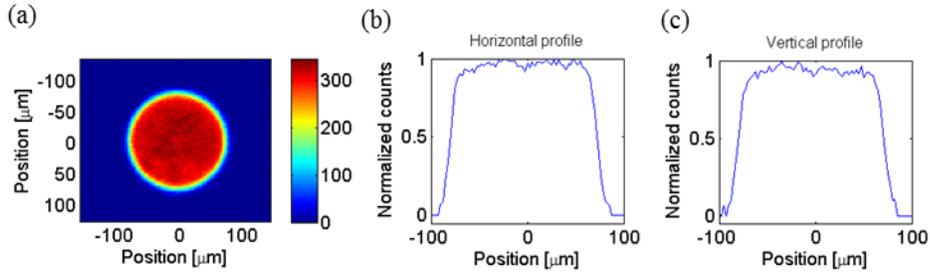

Fig. 5. (a) Near field energy distribution of the launch field at the point of focus (input plane of the chip) for a *F/9* beam and (b-c) its horizontal and vertical line profiles.

## 4. Performance of photonic lanterns

In this section we will summarize results for two sets of photonic lanterns: with varying transition length and varying MM pitch.

*4.1 Transition length*

In order to simplify the characterization process, the structures were composed of MM-SM-MM sections and two transition sections with length varying between 1 and 10 mm. The pitch in the MM section was set to 10 μm. The normalized throughput as a function of the injected *F/#* is shown in Fig. 6. The data show a similar trend as in the simulation results in Section 2.2, i.e. the throughput strongly increases with the increasing transition length up to 5 mm (in simulation) and 3-4 mm (in experiment) and then stabilizes. In the experimental case the sampling set was too small to make detailed comparison with simulation results but nevertheless the overall functional form matches closely. We expect essentially no light to couple in to the guides beyond approximately *F/4*, which is the cutoff *F/#* corresponding to the peak core Δ*n* of $5 \cdot 10^{-3}$. For this reason, we did not make any measurements below *F/4*. The average Δ*n* is $1.3 \cdot 10^{-3}$ which corresponds to the cutoff approximately *F/9*, as shown in our previous work [24]. The increasing throughput between *F/4* and *F/11* is due to the smoothly increasing coupling of higher angles into our complicated multimode guide refractive index profile. The throughput levels off at 0.75. Based on the modeling in section 2.2, the maximum expected throughput (in the case of no phase errors) for a double (MM-SM-MM) transition is 0.94 (0.97 squared). In the case of the presence of phase errors from waveguide inhomogeneities or optical path differences between the waveguides, the main source of loss comes from 19 modes in the SM section into a MM section which only supports 12 modes.

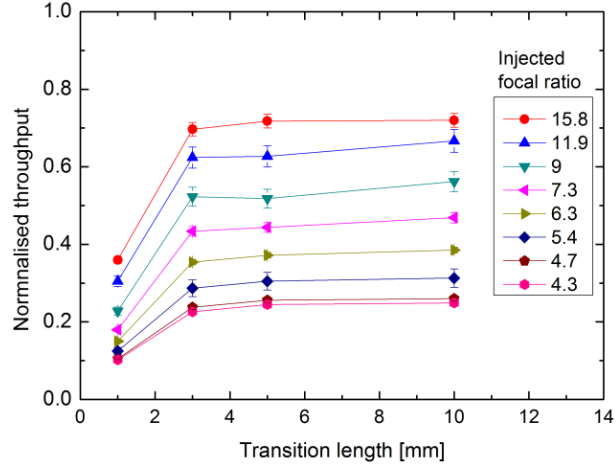

Fig. 6. Back-to-back photonic lantern throughput as a function of transition length for various injection *F/#* probed at 1.55 μm.

*4.2 Waveguide separation in multimode section*

A set of photonic lanterns, composed of MM-SM-MM sections and two transition sections of a fixed length of 10 mm were inscribed with varying pitches in the MM section ranging from 7 to 15 μm in order to investigate how well individual SMs couple back into the MM waveguide. Figure 7 summarizes the normalized throughputs as a function of the injected *F/#*. This figure shows that the throughput increases as a function of pitch until it reaches a peak at 13 μm. Above this pitch the throughput seem to decrease. The reason for the throughput increase up to 13 μm is that, as we showed in the Fig. 1, the 13 um pitch supports almost all modes (17 out of 19). The throughput decreases for structures of a pitch > 13 μm, as the evenly illuminated input light does not couple fully to the spatial profile of the input. In addition we can see that the throughput also increases as an increasing *F/#*. This is predominantly due to the fact that the coupling efficiency is a function of *F/#* and high *F/#* couples better to the structures.

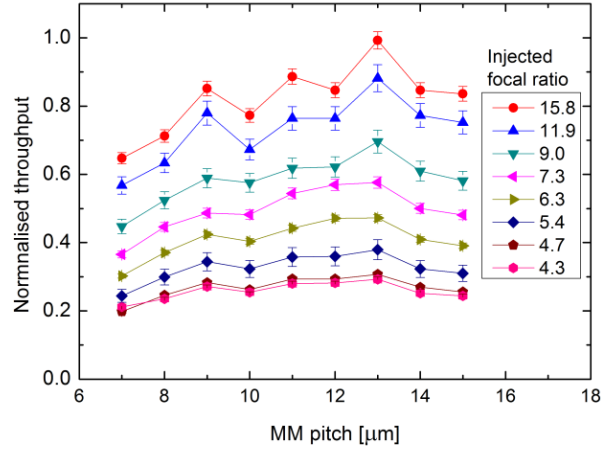

Fig. 7. Back-to-back photonic lantern throughput as a function of MM pitch for various injection *F/#* probed at 1.55 μm.

## 5. Slit reformatting devices for diffraction-limited spectrographs

An ideal application for photonic lantern is to convert the MM light into SM light and then reformat it into a slit to be directly injected into a spectrograph [29,30,31,23]. The slit reformatting device, which consists of a MM input, a transition section and isolated SM tracks which are remapped to a linear slit of SM waveguides is shown in Fig. 8(a). Figure 8(b) shows a concept illustration of spectra produced from a multiple SM-waveguide slit as viewed on a CCD detector. The figure depicts the architecture for optimum detector sampling.

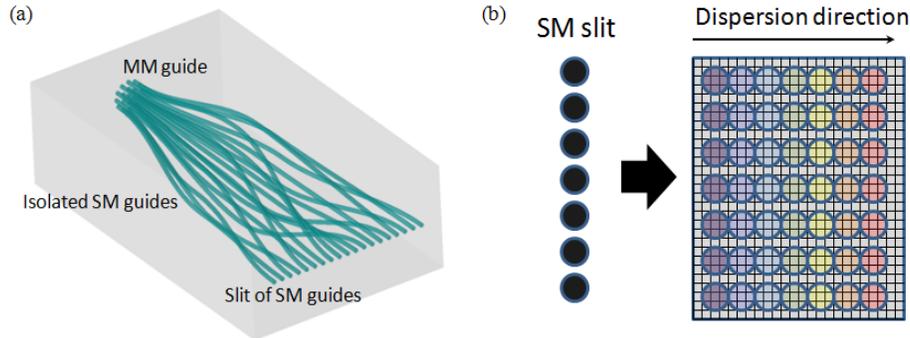

Fig. 8. (a) A CAD model of a slit reformatting device for a diffraction-limited spectrograph. Three key regions are highlighted; the MM guide that collects the seeing-limited light from the focus of the telescope, the isolated SM guides and the slit formed from the SM guides. (b) Concept illustration of spectra viewed on a CCD detector produced by a long, thin, SM slit. The required sampling is at least 2.4 pixels per resolution element (here 3 pixels). A SM-waveguide slit allows for optimum use of the detector area.

*5.1 The number of modes supported by the device*

Once the MM waveguides and transition type are optimized, the next major consideration is to ensure that the number of modes supported by each section of the slit reformatting device is the same. As shown in section 2.2 an efficient transition between the MM and SM sections can be achieved when the number of non-degenerate modes in the MM section is less than or equal to the number of SM tracks (19). The now separated SM beams could be reformatted into a slit with separations between SM tracks sufficient to avoid coupling. This however would require a long slit (19 guides at the wavelength of 1.55 μm, 40 μm spacing means 720 μm slit length) which becomes difficult when one moves to the visible as the mode count scales with $1/\lambda^2$ (150 guides now required at the wavelength of 0.55 μm, 15 μm spacing means 2250 μm slit length). This puts constraints on the quality of the optics required to image off-axis cores at the diffraction limit in the spectrograph. A more serious concern though is the poor use of the valuable detector area.

As such the ideal scenario is to bring the waveguides in the slit closer together so that they become coupled. However, when recombining the beams to form a slit, the converse of the above condition is also true; for a lossless slit the number of non-degenerate modes supported by the slit must be equal to the number of SM beams that are being combined. This condition can be fulfilled by careful selection of the waveguide spacing in the slit. The number of modes supported by a linear array of SM tracks as a function of the spacing has been calculated in FemSIM and is also shown in Fig. 1 in Section 2.1 (as for comparison the results were overlaid with the number of modes for a circular arrangement). It can be seen that a minimum separation of 13 μm between the tracks will still support the 19 modes required for a lossless slit to be realized. By using this technique it is possible to reduce the slit length to 250 μm and hence maximize the detector real estate used. In this way the slit is MM in nature along its length but still retains the SM format in the orthogonal dispersion direction, which allows for a miniature spectrograph (free from scaling laws common to seeing-limited spectrographs) to be constructed.

*5.2 Performance of the slit reformatting devices.*

A set of slit reformatting devices was inscribed with varied separation between the SM waveguides in the slit end, i.e. slit pitches ranging from 7-17 μm. The input MM pitch was set to 10 μm and the length of two transition sections was set to 10 mm. The structures were characterized as described in Section 3. The normalized throughputs of the structures as a function of the injected *F/#* and the slit pitch are presented in Fig. 9. The maximum normalized throughput is obtained for slit pitches $\geq 13$ μm. This coincides with the theoretical value discussed above (Fig. 1, Section 2.1). Most notably it can be seen for slow beams (high *F/#*) that have fewer spatial modes, the normalized throughput is ~100% to within uncertainty at 1.55 μm. This means that aside from the coupling and absorption losses the MM to SM conversion and reformatting is near lossless.

The high efficiency of devices with 13 μm separation between tracks in the slit means that the total slit size would be ~12 μm (mode field diameter) × 240 μm (slit length), in contrast to the similar device based on standard SMF 28 fibers, which would be ~10.5 μm (mode field diameter) × 2400 μm (slit length), so a factor of 10 along the slit. As a result, the higher density of the waveguides will allow for better utilization of the detector area.

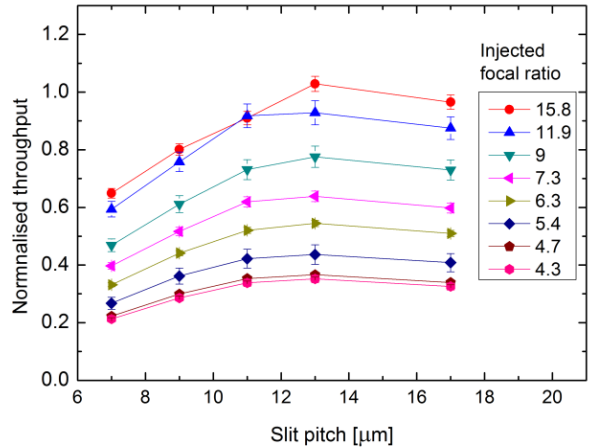

Fig. 9. Normalized throughput for the slit reformatting device for various slit pitches (separation between SM waveguides in the slit end) as a function of the injected focal ratio.

## Conclusion

We have evaluated the design and fabricated a series of photonic lanterns and slit reformatting devices in order to find a recipe for low-loss devices. The simulations showed that out of linear, cosine and raised sine bend types the cosine one has the lowest losses. The throughput measurement showed that devices with a MM-SM transition as short as 3-5 mm are practical and exhibit efficiencies at the 75% level. The study of the pitch of the MM section demonstrated that it is critical to optimize the pitch for efficient coupling. In our case this corresponded to a pitch of ≥13 μm which supported all the modes. In addition we minimized the length of the slit in the reformatting device by using a 13 μm pitch which made the slit multimode and maintained the throughput of 95% for F/#>12 (excluding glass absorption losses).

The promising performance of the integrated photonic lanterns and slit reformatting devices opens up the possibility of using them to deliver the light from the telescope focus to the single-mode spectrographs, either the integrated photonic spectrographs or the bulk optics-based single mode spectrographs. Now that the efficient photonic lanterns can be realized in the integrated platform astrophotonic devices can be greatly miniaturized and hence made more thermally and environmentally stable. Such instrument can greatly enhance the efficiency and performance of the seeing-limited telescopes.

## Acknowledgments

We would like to thank Dr Sergio Leon-Saval for fruitful discussions in regards to photonic lanterns. This research was supported by the Australian Research Council Centre of Excellence for Ultrahighbandwidth Devices for Optical Systems (project no. CE110001018) and the OptoFab node of the Australian National Fabrication Facility. I. Spaleniak acknowledges the support of the iMQRES scholarship and AAO top-up scholarship.